\documentclass[a4paper]{article}

\usepackage{INTERSPEECH2021}
\usepackage{graphicx,xcolor}
\usepackage{slashbox}
\usepackage{multirow}

\title{End-to-End Speaker-Attributed ASR with Transformer}

\name{Naoyuki Kanda, Guoli Ye, Yashesh Gaur, Xiaofei Wang, Zhong Meng, \\Zhuo Chen, Takuya Yoshioka}
\address{
  Microsoft Corp., USA} 
\email{\{Naoyuki.Kanda,guoye,Yashesh.Gaur,Xiaofei.Wang,Zhong.Meng,zhuc,tayoshio\}@microsoft.com}

\begin{document}

\maketitle
\begin{abstract}
This paper presents our recent effort on end-to-end  speaker-attributed automatic speech recognition,
which jointly performs speaker counting, speech recognition and speaker identification for monaural multi-talker audio.
Firstly, we thoroughly update the model architecture that was previously designed based on a long short-term memory (LSTM)-based attention encoder decoder by applying transformer architectures.
Secondly, we 
propose a speaker deduplication mechanism to 
reduce speaker identification errors in highly overlapped regions. 
Experimental results on the LibriSpeechMix dataset
shows that the transformer-based architecture is especially good at counting the speakers 
and that 
the proposed model reduces the speaker-attributed word error rate by 47\% over the LSTM-based baseline. 
Furthermore, for the LibriCSS dataset, which consists of real recordings of overlapped speech,  
the proposed model achieves concatenated minimum-permutation word error rates of 11.9\% and 16.3\%
with and without target speaker profiles, respectively, 
both of which
are the state-of-the-art results 
for LibriCSS with the monaural setting.
\end{abstract}
\noindent\textbf{Index Terms}: multi-speaker speech recognition,
speaker counting, speaker identification, 
serialized output training

\section{Introduction}

Speaker attributed automatic speech recognition (SA-ASR) 
for recognizing ``who spoke what'' from multi-talker recordings has long been 
studied to enable conversation analysis
 \cite{janin2003icsi,carletta2005ami,fiscus2007rich}.
It requires counting the participating speakers, transcribing each speaker's utterance,
and assigning a speaker tag to each utterance from possibly overlapped speech.
One approach for SA-ASR is to first 
 perform speech separation and then apply 
ASR and speaker diarization/identification to the separated audio streams (e.g., \cite{yoshioka2019advances,raj2020integration}).
However, 
 such combination could limit the overall accuracy
since each module is separately trained with its own criterion. 
To overcome this suboptimality, 
a joint modeling approach has been studied widely, ranging from 
partially joint modeling methods (speech separation plus ASR \cite{yu2017recognizing,seki2018purely,kanda2019acoustic,kanda2019auxiliary,sklyar2020streaming,lu2020streaming}, speaker identification plus speech separation methods \cite{wang2019speech,von2019all},
etc.) to fully joint modeling \cite{el2019joint,mao2020speech,kanda2019simultaneous}.

Recently, an end-to-end (E2E) SA-ASR model was proposed to jointly perform
all the sub-tasks of SA-ASR regardless of the number of speakers in an input recording \cite{kanda2020joint,kanda2020minimum}.
The E2E SA-ASR consists of an ASR-block and a speaker-block. 
Both blocks
were designed by using 
attention-based encoder decoders (AEDs) based on long short-term memory (LSTM), and the two blocks worked interdependently
to perform ASR and speaker identification.
The E2E SA-ASR model showed significant improvement in speaker-attributed word error rate (SA-WER) compared with a system combining multi-talker ASR 
and speaker identification.
Furthermore, it was also shown that the E2E SA-ASR could
be combined with speaker clustering 
to perform speaker diarization when relevant speaker profiles were unavailable~\cite{kanda2020investigation}.
However, 
noticeable degradation of accuracy was still observed especially 
when the number of overlapping speakers became large.

In this paper, 
we present our recent progress on further improving the E2E SA-ASR.
We first
apply the recent advancement of transformer architecture \cite{vaswani2017attention,dong2018speech,zhou2018syllable,karita2019comparative,zeyer2019comparison,gulati2020conformer} to
both the ASR- and speaker-blocks of the E2E SA-ASR. 
We also propose a simple modification to the joint inference procedure, 
named speaker deduplication, to
prevent the system from predicting the same speaker in highly overlapped regions.
Evaluation on two datasets shows that
the proposed model is especially good at counting the number of speakers
and it can significantly improve the accuracy
over the LSTM-based baseline.

\section{E2E SA-ASR: Review}
\subsection{Problem statement}

Suppose 
we observe a sequence of acoustic feature $X\in\mathbb{R}^{f^a\times l^a}$, 
where $f^a$ and $l^a$ are the feature dimension and the sequence length,
respectively.
Also suppose 
that a set of 
 speaker profiles $D=\{d_k \in \mathbb{R}^{f^d}|k=1,...,K\}$ is available,
where $K$ is the total number of the profiles, 
 $d_k$ is a speaker embedding 
 (e.g., d-vector \cite{variani2014deep})
of the $k$-th speaker,
and $f^d$ is the dimension of the speaker embedding.
$K$ can be greater than the actual number of speakers in the recording.

The goal of E2E SA-ASR \cite{kanda2020joint}
  is 
  to estimate a
  multi-speaker transcription $Y=(y_n\in \{1,...,|\mathcal{V}|\}|n=1,...,N)$
accompanied by
the speaker identity of each token $S=(s_n\in \{1,...,K\}|n=1,...,N)$ 
 given
 input 
 $X$ and $D$.
Here,
$|\mathcal{V}|$  is the size of the vocabulary $\mathcal{V}$, 
and  $y_n$ and $s_n$ are the word index and speaker index for the $n$-th token,
 respectively.
  Following the serialized output training (SOT) framework \cite{kanda2020sot}, we represent 
a multi-speaker transcription as follows: 
word sequences of multiple speakers are joined 
 by a special change symbol $\langle sc\rangle$ to form a single sequence $Y$.
 For example, for the three-speaker case, the reference token sequence to $Y$ is given as
$R=\{r^1_{1},..,r^1_{N^1}, \langle sc\rangle, r^2_{1},..,r^2_{N^2}, \langle sc\rangle, r^3_{1},..,r^3_{N^3}, \langle eos\rangle\}$, 
where $r^j_i$ represents the $i$-th token of the $j$-th utterance.
 
\subsection{Model architecture}

\begin{figure}[t]
  \centering
  \includegraphics[width=0.95\linewidth]{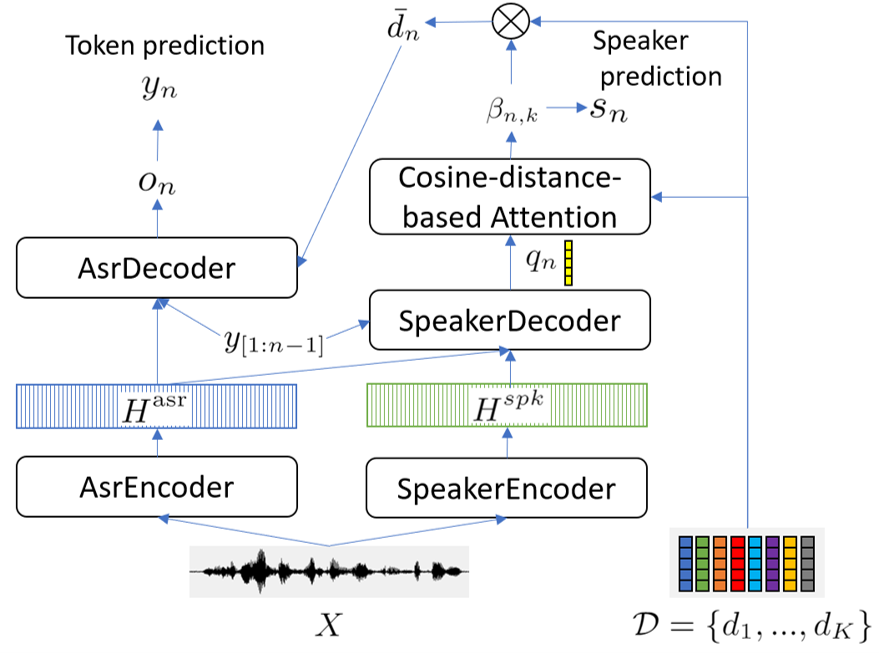}
  \vspace{-3mm}
      \caption{Overview of E2E SA-ASR}
  \label{fig:asr_models}
  \vspace{-5mm}
\end{figure}

The E2E SA-ASR model consists of 
the {\it ASR block} and the {\it speaker block},
which jointly perform
ASR and speaker identification (Fig. \ref{fig:asr_models}). 
The ASR block follows an encoder-decoder design and is 
represented as, 
 \begin{align}
 H^{\rm asr} &={\rm AsrEncoder}(X),  \label{eq:enc}  \\
 o_n &= {\rm AsrDecoder}(y_{[1:n-1]}, H^{\rm asr}, \bar{d}_n).  \label{eq:asrout}
 \end{align}
Given the acoustic input $X$, 
an AsrEncoder module, which will be detailed in the next section, 
first converts $X$ 
into a sequence of hidden embeddings $H^{\rm asr} \in \mathbb{R}^{f^h\times l^h}$ for ASR (Eq. \eqref{eq:enc}),
where $f^h$ and $l^h$ are the embedding dimension and 
the length of the sequence, respectively.
At each decoder step $n$, 
AsrDecoder module 
calculates
the output distribution $o_n \in \mathbb{R}^{|\mathcal{V}|}$ 
 given
previous token estimates $y_{[1:n-1]}$,
$H^{\rm asr}$,
and 
the weighted speaker profile $\bar{d}_n$ (Eq. \eqref{eq:asrout}).
Note that $\bar{d}_n$ is computed 
from the speaker profiles $D$
in the speaker block, which will be
explained later.
The posterior probability
of token $i$ (i.e. the $i$-th token in $\mathcal{V}$) 
at the $n$-th decoder step 
is represented as
 \begin{align}
Pr(y_n=i|y_{[1:n-1]},s_{[1:n]},X,D) = o_{n,i}, \label{eq:tokenprob}
\end{align}
where $o_{n,i}$ represents
the $i$-th element of $o_n$.

On the other hand, the speaker block is represented as 
\begin{align}
 H^{spk} &= {\rm SpeakerEncoder}(X),  \label{eq:spkenc} \\
 q_n &= {\rm SpeakerDecoder}(y_{[1:n-1]},H^{\rm spk},H^{\rm asr}), \label{eq:spkquery} \\
\beta_{n,k}&= \frac{\exp(\cos(q_n,d_k))}{\sum_j^K \exp(\cos(q_n,d_j))}, \label{eq:invatt} \\
\bar{d}_n&=\sum_{k=1}^{K}\beta_{n,k}d_k. \label{eq:weighted_prof}
\end{align}
The SpeakerEncoder module first converts $X$ into 
a speaker embeddings $H^{\rm spk}\in \mathbb{R}^{f^h \times l^h}$
that represents the speaker characteristic of
 $X$ (Eq. \eqref{eq:spkenc}).
At every decoder step $n$,
SpeakerDecoder
calculates
a speaker query $q_n \in \mathbb{R}^{f^d}$ given $y_{[1:n-1]}$,
$H^{\rm spk}$ and $H^{\rm asr}$ (Eq. \eqref{eq:spkquery}).
Next,
a cosine distance-based attention weight
 $\beta_{n,k}\in \mathbb{R}$
is calculated 
for all 
 profiles $d_k$ in $D$
given the speaker query $q_n$ (Eq. \eqref{eq:invatt}).
The $\beta_{n,k}$ can be seen as 
a posterior probability of person $k$ speaking the $n$-th token
given all the previous estimation as well as $X$ and $D$, i.e., 
\begin{align}
Pr(s_n=k|y_{[1:n-1]},s_{[1:n-1]},X,D)=\beta_{n,k}. \label{eq:spk-prob}
\end{align}
Finally,
 the weighted speaker profile $\bar{d}_n\in \mathbb{R}^{f^d}$
 is calculated as the weighted average of the profiles in $D$
 (Eq. \eqref{eq:weighted_prof}),
 which 
 is fed to the ASR block 
 (Eq. \eqref{eq:asrout}).

By using Eqs. \eqref{eq:tokenprob} and \eqref{eq:spk-prob}, 
the joint posterior probability of token $Y$ and speaker $S$ given input $X$ and $D$
is represented as, 
\begin{align}
Pr(Y,S|X,D) =&\prod_{n=1}^{N}\{Pr(y_{n}|y_{[1:n-1]}, s_{[1:n]}, X, D) \nonumber \\ 
&\;\;\times Pr(s_{n}|y_{[1:n-1]}, s_{[1:n-1]}, X, D) \}. \label{eq:samll-2}
\end{align}
The model parameters are optimized by maximizing $Pr(Y,S|X,D)$ over
training data.

\section{Proposed method}
\subsection{Transformer-based SA-ASR architecture}
\label{sec:detail}

In the prior work \cite{kanda2020joint}, AsrEncoder, AsrDecoder, SpeakerDecoder were represented 
by the stack of LSTM layers.
In this paper, we  thoroughly  update  the  model  architecture
by applying the recent advancement of the transformer.
Note that 
the AsrDecoder and SpeakerDecoder  requires
 a specical design to work jointly,
 so we use most of the space to explain them.

\underline{\bf AsrEncoder} is represented by a modified version of Conformer network \cite{gulati2020conformer}.
We follow the architecture shown in \cite{gulati2020conformer}
except 
 that
 (i) we insert an squeeze and excitation module \cite{hu2018squeeze} 
  before the dropout of the convolution module
and (ii) we add one more point-wise convolution after 
depth-wise convolution. 
These changes were made based on our preliminary test.

\underline{\bf AsrDecoder} is similar to a conventional 
transformer-based decoder except
that the weighted speaker profile $\bar{d}_n$ 
is added
at the first layer.
The AsrDecoder is represented as follows.
\begin{align}
&z_{[1:n-1],0}^{\rm asr}=\mathrm{PosEnc}(\mathrm{Embed}(y_{[1:n-1]})),\label{eq:emb}\\
&\bar{z}^{\rm asr}_{n-1,l}=z^{\rm asr}_{n-1,l-1}  \nonumber \\
& \hspace{3mm}+ \mathrm{MHA}_{\rm l}^{\rm asr{\text -}self}(z^{\rm asr}_{n-1,l-1},z_{[1:n-1],l-1}^{\rm asr},z_{[1:n-1],l-1}^{\rm asr}), \label{eq:asr-self}\\
&\bar{\bar{z}}^{\rm asr}_{n-1,l}=\bar{z}^{\rm asr}_{n-1,l} + \mathrm{MHA}_{\rm l}^{\rm asr{\text -}src}(\bar{z}^{\rm asr}_{n-1,l},H^{\rm asr},H^{\rm asr}), \label{eq:asr-src}\\
&z^{\rm asr}_{n-1,l+1}=
\left\{
\hspace{-2mm}\begin{array}{ll}
\bar{\bar{z}}^{\rm asr}_{n-1,l}+\mathrm{FF}_l^{\rm asr}(\bar{\bar{z}}^{\rm asr}_{n-1,l} + W^{\rm spk}\cdot \bar{d}_n) & \hspace{-3mm}(l=1)  \\
\bar{\bar{z}}^{\rm asr}_{n-1,l}+\mathrm{FF}_l^{\rm asr}(\bar{\bar{z}}^{\rm asr}_{n-1,l})& \hspace{-3mm} (l>1) 
\end{array}
\right. \label{eq:asr-ff}\\
&o_{n} = \mathrm{SoftMax}(W^{o}\cdot z_{n-1,L^{\rm asr}}^{\rm asr} + b^o) \label{eq:asr_out}
\end{align}
Here, $\mathrm{Embed}()$ and $\mathrm{PosEnc}()$ are the embedding
function and absolute positional encoding function \cite{vaswani2017attention}, respectively.
$\mathrm{MHA}^*_l(Q,K,V)$ represents the multi-head attention of the $l$-th layer
\cite{vaswani2017attention} with query $Q$, key $K$, and value $V$ matrices.
$\rm FF_l^*()$ represents a position-wise feed forward network of $l$-th layer.
In the AsrDecoder, token sequence $y_{[1:n-1]}$ is first converted
into a sequence of embedding $z_{[1:n-1],0}^{\rm asr}\in \mathbb{R}^{f^h\times (n-1)}$
(Eq. \eqref{eq:emb}).
For each layer $l$ in the AsrDecoder,
 the self-attention operation 
(Eq. \eqref{eq:asr-self}) 
 and source-target attention operation (Eq. \eqref{eq:asr-src}) are 
 applied.
 Finally, the position-wise feed forward layer is applied to
 calculate the input to the next layer 
$z_{n,l+1}^{\rm asr}$ (Eq. \eqref{eq:asr-ff}).
Here, unlike the conventional transformer-based decoder,
$\bar{d}_n$ is added after being multiplied by the weight $W^{spk} \in \mathbb{R}^{f^h \times f^d}$
in the first layer.
Finally, $o_n$
is calculated by applying SoftMax function 
on the final $L^{\rm asr}$-th  layer's output
with
weight $W^o \in \mathbb{R}^{|\mathcal{V}| \times f^h}$ and bias $b^o \in \mathbb{R}^{|\mathcal{V}|}$ 
application
 (Eq. \eqref{eq:asr_out}). 

\begin{table*}[t]
\setcounter{table}{1}
  \caption{SER (\%), WER (\%), and SA-WER (\%) for LibriSpeechMix with
baseline and proposed models. LM was not used.}
  \label{tab:baseline}
  \vspace{-3mm}
  \centering
{\scriptsize
  \begin{tabular}{cl|ccc|ccc|ccc|ccc}
    \toprule
Setting &\multirow{2}{*}{\backslashbox{Model}{Eval Set}} &  \multicolumn{3}{c|}{1-speaker} & \multicolumn{3}{c|}{2-speaker-mixed } & \multicolumn{3}{c|}{3-speaker-mixed} & \multicolumn{3}{c}{\bf{Total}}\\
ID& &  SER & WER & {\bf SA-WER} &  SER & WER & {\bf SA-WER} &  SER & WER & {\bf SA-WER} &   SER &  WER & {\bf SA-WER}  \\
    \midrule
&LSTM SA-ASR \cite{kanda2020joint} & 0.2 & 4.2 & {\bf 4.5}  & 2.5 & 8.7 & {\bf 9.9} & 10.2 & 20.2 & {\bf 23.1} &  6.0 &  13.7 & {\bf 15.6} \\ \midrule 
(a)&Transformer SA-ASR & 0.5 & 3.3 & {\bf 3.8}  & 2.1 & 5.3 & {\bf 6.5} & 4.9 & 8.9 & {\bf 10.7} & 3.2 & 6.8  & {\bf 8.2} \\ 
(b)& \hspace{3mm}+ SpecAugment  & 0.5 & 3.2 & {\bf 3.8}  & 2.3 & 5.3 & {\bf 6.8}  & 4.6 & 7.7 & {\bf 9.4} & 3.1 & 6.2   & {\bf 7.6} \\ 
(c)&\hspace{6mm}+ Speaker Dedupulication &  0.6 & 3.3 & {\bf 3.9} & 2.0 & 4.3 & {\bf 6.4} & 3.3 & 6.0  & {\bf 8.5} & 2.4 & 5.0  & {\bf 7.0} \\ 
     \bottomrule
  \end{tabular}
}
  \vspace{-5mm}
\end{table*}

\underline{\bf SpeakerEncoder} is almost identical 
to the speaker profile extractor (in our case, a d-vector extractor 
based on a convolutional neural network \cite{zhou2019cnn})
except that the final average pooling layer in the speaker profile extractor is 
replaced by a linear layer without pooling.
The last linear layer maps $f^d$-dim output of the d-vector 
extractor into $f^h$-dim embeddings.

\underline{\bf SpeakerDecoder} is represented by a similar architecture to the transformer-based decoder,
but it is designed to share the source-target attention information with that of the AsrDecoder.
The first layer of the SpeakerDecoder is represented as
\begin{align}
\bar{\bar{z}}^{\rm spk}_{n-1,1}&=\bar{z}^{\rm asr}_{n-1,1} + \mathrm{MHA}_{n-1,1}^{\rm spk{\text -} src}(\bar{z}^{\rm asr}_{n-1,1},H^{\rm asr},H^{\rm spk}), \label{eq:spk-src1}\\
z^{\rm spk}_{n-1,2}&= \bar{\bar{z}}^{\rm spk}_{n-1,1}+\mathrm{FF}_1^{\rm spk}(\bar{\bar{z}}^{spk}_{n-1,1}). \label{eq:spk-ff1}
\end{align}
Here, we reuse $\bar{z}_{n-1,1}^{\rm asr}$,
derived from $y_{[1:n-1]}$ in the ASR block,
as the query to the key of
$H^{\rm asr}$ (Eq. \eqref{eq:spk-src1}).
The value is calculated based on $H^{\rm spk}$ so that the speaker representation
is obtained as $\bar{\bar{z}}_{n-1,1}^{\rm spk}\in \mathbb{R}^{f^h}$.
Point-wise feed forward layer is then applied to obtain
 $z^{\rm spk}_{n-1,2} \in \mathbb{R}^{f^h}$ (Eq. \ref{eq:spk-ff1}).
The architecture after the first layer is 
represented as follows.
\begin{align}
&\bar{z}^{\rm spk}_{n-1,l}=z^{\rm spk}_{n-1,l-1} \nonumber \\
&\hspace{1mm}+ \mathrm{MHA}_{\rm l}^{\rm spk{\text -}self}(z^{\rm spk}_{n-1,l-1},z^{\rm spk}_{[1:n-1],l-1},z^{\rm spk}_{[1:n-1],l-1}), \label{eq:spk-self-2}\\
&\bar{\bar{z}}^{\rm spk}_{n-1,l-1}=\bar{z}^{\rm spk}_{n-1,l} + \mathrm{MHA}_{\rm l}^{\rm spk{\text -}src}(\bar{z}^{\rm spk}_{n-1,l},H^{\rm spk},H^{\rm spk}), \label{eq:spk-src-2}\\
&z^{\rm spk}_{n-1,l+1}= \bar{\bar{z}}^{\rm spk}_{n-1,l}+\mathrm{FF}_l^{\rm spk}(\bar{\bar{z}}^{\rm spk}_{n-1,l}) \label{eq:spk-ff}\\
&q_n =W^{q}\cdot z^{\rm spk}_{n-1,L^{\rm spk}} \label{eq:spk-out}
\end{align}
Here, the self attention, source-target attention, and point-wise feed forward layer
are applied  as with the conventional transformer decoder (Eqs. \eqref{eq:spk-self-2}--\eqref{eq:spk-ff}).
Finally, the speaker query $q_n$ is obtained by multiplying a weight $W^q \in \mathbb{R}^{f^d \times f^h}$
to the final $L^{\rm spk}$-th layer's output
(Eq. \eqref{eq:spk-out}).

\subsection{Inference with speaker deduplication}
At inference time, the extended beam search algorithm \cite{kanda2020joint} is used
to generate transcriptions with speaker labels.
In the prior work, 
the speaker with the highest average $\beta_{n,k}$ score for each utterance is 
simply selected as the predicted speaker of that utterance.
However, it was found that 
the same speaker was sometimes predicted for heavily overlapped utterances of different speakers. 
To avoid this kind of errors,
we set a simple constraint that 
the same speaker cannot be assigned for
the consecutive utterances joined by $\langle sc\rangle$.
Given that constraint, a sequence of speakers that has the highest 
speaker probability (multiplication of $\beta_{n,k}$) among all possible speaker sequence is selected.
We call this method as ``speaker deduplication" in this paper.

\begin{table}[t]
\setcounter{table}{0}
  \caption{WER (\%) comparison of ASR-block for LibriSpeechMix. LM was not used. }
  \label{tab:sot}
  \vspace{-3mm}
  \centering
{\scriptsize
  \begin{tabular}{lc|ccc|c}
    \toprule
 Model & \# of & \multicolumn{3}{c|}{{\scriptsize\# of Speakers in Test Data}}& Total \\
    &   Params.    & 1 & 2 & 3 &   \\
    \midrule
SOT LSTM-AED \cite{kanda2020joint} &  135.6M &  4.5 &  10.3 &  19.5 & 13.9  \\ \midrule
SOT Transformer-AED      &  128.6M & 4.1 & 5.3 & 7.5 & 6.2 \\ 
  \hspace{3mm}+ SpecAugment      &  128.6M & 3.6 & 4.9 & 6.2 & 5.3\\ 
    \bottomrule
  \end{tabular}
  }
  \vspace{-5mm}
\end{table}

\section{Experiments}
We first conducted experiments based on LibriSpeechMix \cite{kanda2020joint}, which is 
a clean mixed-audio test set,
to assess
the basic capability of the proposed model.
We then conducted experiments based on LibriCSS \cite{chen2020continuous}, which is 
a set of multi-talker long-form audio
recorded in a real meeting room. 

\subsection{Evaluation with LibriSpeechMix}
\subsubsection{Evaluation data and evaluation metric}

 LibriSpeechMix 
 includes single-speaker data, two-speaker-mixed data, and three-speaker mixed data,
 all of which were made by mixing utterances of ``test\_clean" of LibriSpeech \cite{panayotov2015librispeech} 
 without changing the signal level of each utterance.
Each utterance was randomly delayed to simulate partially overlapped speech as seen in natural conversations.
For each test sample, 
8 speaker profiles, or 128-dim d-vectors \cite{zhou2019cnn}, 
are extracted for the speaker identification task.
The utterances to be used for speaker profile extraction are determined in the dataset.
Speaker error rate (SER), WER, and SA-WER as defined in \cite{kanda2020joint} were used
for the evaluation.

\begin{table}[t]
  \setcounter{table}{2}
  \caption{Speaker counting accuracy (\%) for LibriSpeechMix.}
  \label{tab:spk-count}
  \vspace{-3mm}
  \centering
  {\scriptsize
  \begin{tabular}{l|c|cccc}
    \toprule
Model & Actual \# of Speakers& \multicolumn{4}{c}{Estimated \# of Speakers (\%)} \\
      & in Test Data & 1 & 2 & 3 & $>$4 \\ \midrule
LSTM                    & 1 & {\bf 99.96} & 0.04 & 0.00 & 0.00\\ 
SA-ASR \cite{kanda2020joint} & 2 & 2.56 & {\bf 97.44} & 0.00 & 0.00 \\ 
                    & 3 & 0.31 & 25.34 & {\bf 74.35} & 0.00\\ 
    \midrule
 Transformer & 1 & {\bf 100.00} & 0.00 & 0.00 & 0.00 \\ 
  SA-ASR & 2 & 2.21 & {\bf 97.79 } & 0.00 & 0.00 \\ 
(setting (a))  & 3 & 0.31 & 10.15 & {\bf 89.54} & 0.00\\ \midrule
 Transformer & 1 & {\bf 99.92} & 0.08 & 0.00 & 0.00 \\ 
 SA-ASR & 2 & 0.53 & {\bf 99.39 } & 0.08 & 0.00 \\ 
(setting (c)) & 3 & 0.00 & 3.74 & {\bf 96.26} & 0.00\\ 
    \bottomrule
  \end{tabular}
  }
  \vspace{-5mm}
\end{table}

\begin{table*}[t]
  \caption{Comparison of cpWER (\%) for LibriCSS with various methods.}
  \label{tab:auto_summary}
  \vspace{-3mm}
  \centering
{\scriptsize
\begin{tabular}{lccc|cccccc|c}
    \toprule
System &\# of& Speaker & Speaker&  \multicolumn{7}{c}{cpWER (\%) for different overlap ratio} \\
&channel& Profile& Counting &  0S & 0L & 10 & 20 & 30 & 40 & {\bf Avg.}  \\ \midrule
7ch-CSS + SC + Transformer-ASR \cite{raj2020integration}& 7 & - & automatic & 12.5 & 9.6 & 12.7 & 12.9 & 14.4 & 13.5 & {\bf 12.7} \\ 
TS-VAD + Transformer-ASR \cite{raj2020integration}& 1 & -  & oracle &11.0 & 9.5 & 16.1 & 23.1 & 33.8 & 40.9 & {\bf 23.9} \\ \midrule
LSTM SA-ASR \cite{kanda2020investigation} & 1 &  $\surd$ & automatic &15.7  & 8.0 & 12.5 & 17.5 & 24.3 & 27.6 & {\bf 18.6}  \\  
 LSTM-SA-ASR + SC \cite{kanda2020investigation} & 1 & - & oracle & 15.8 & 10.3 & 13.4  & 17.1 & 24.4 & 28.6 & {\bf 19.2} \\ 
 LSTM SA-ASR + SC \cite{kanda2020investigation} & 1 &  - & automatic     & 24.4  & 12.2   & 15.0 & 17.1  & 28.6 & 28.6 & {\bf 21.8}  \\ \midrule
Transformer SA-ASR & 1 & $\surd$  & automatic &  12.1  & 7.9 & 9.6  & 10.8 & 13.4  & 15.7 & {\bf 11.9}  \\  
Transformer SA-ASR + SC & 1 & - & oracle & 12.7 & 8.6 & 11.2  & 11.3  & 16.1  & 17.5 & {\bf 13.3}  \\  
Transformer SA-ASR + SC & 1 & - & automatic & 14.7 &  10.4 & 16.3 & 14.7 & 21.3 & 17.5 & {\bf 16.3}  \\  \bottomrule
  \end{tabular}
  }
\\{\footnotesize  CSS: Continuous Speech Separation, SC: Spectral Clustering, TS-VAD: Target-speaker Voice Activity Detection}
  \vspace{-5mm}
\end{table*}

\subsubsection{Model and training settings}

We used a 80-dim log mel filterbank extracted every 10 msec as the input feature.
For the speaker profile, we used
a 128-dim d-vector \cite{variani2014deep}, 
whose extractor was separately 
trained on VoxCeleb Corpus \cite{nagrani2017voxceleb,chung2018voxceleb2}.
Our d-vector extractor 
consisted of 17 convolution layers 
followed by an average pooling layer, 
which was a modified version of the one presented in \cite{zhou2019cnn}. 

The AsrEncoder consisted of 
2 layers of convolution layers that subsamples the time frame by a factor of 4,
followed by
18 conformer layers.
Each conformer layer consisted of
two 1024-dim feed forward layers in a sandwich structure,
a multi-head attention with 8 heads,
a depthwise convolution with kernel size 3,
and a squeeze-and-excitation network with reduction factor 8
\cite{hu2018squeeze}.
Embedding dimension $f^h$ was set to 512.
The AsrDecoder consisted of
6 layers, each of which had the
multi-head attention with 8 heads, and 2048-dim
feed forward layer.
16k subwords \cite{kudo2018subword}
were used as a recognition unit.
The SpeakerEncoder was the same 
as the d-vector extractor except 
the final layer
 as explained in Section \ref{sec:detail},
 and initialized by the parameter of the d-vector extractor.
 Finally, SpeakerDecoder consisted of 2 layers,
 each of which had a multi-head attention with
 8 heads and a 2048-dim feed forward layer.

Following
 the previous work \cite{kanda2020joint}, 
we first optimized only the ASR block as a speaker-agnostic multi-speaker ASR model
by setting $\bar{d}_n=0$.
The training data was the same as the one used in \cite{kanda2020sot}.
Namely, we generated multi-speaker audio segments by randomly mixing 1 to 3 utterances of LibriSpeech ``train\_960" 
with a random delay for each utterance, where the minimum difference of the starting time of each utterance was set to 0.5 sec. 
We used a mini-batch of 12,000 frames and trained the model for 160k iterations with 32 GPUs, with 
Noam learning rate schedule with a peak learning rate of 0.0002 after 25k iterations.
After the 160k iterations, we reset the optimizer and conducted the
additional 320k iterations of training with SpecAugment \cite{park2019specaugment}.
Table \ref{tab:sot} shows the comparison of the WER by the transformer-based model
(at the first 160k iteration w/o SpecAugment, and after additional 320k iteration with SpecAugment)
and the LSTM-based model \cite{kanda2020joint}.
We can observe a significant WER improvement especially for the 2-speaker and 3-speaker mixed test data.

Given the well-trained ASR-block as the initial parameters,
we further trained the entire model parameters (i.e. including the speaker block)
by the training data used in \cite{kanda2020joint}.
The training data were again generated from LibriSpeech by randomly mixing 1 to 3 utterances 
 with a minimum delay of  0.5 sec.
We used a mini-batch of 6,000 frames and trained 160k iterations with 8 GPUs, with 
Noam learning rate schedule with a peak learning rate of 0.0001 after 10k iterations.
We did not use SpecAugment at this stage since it degraded the accuracy.

\subsubsection{Evaluation results}
Table \ref{tab:baseline} shows the comparison of the LSTM-based SA-ASR  and the proposed transformer-based SA-ASR.
Note that the setting (a) is the SA-ASR model initialized by the ASR-block w/o SpecAugment
(middle row of Table \ref{tab:sot}) while the setting (b) is the SA-ASR model
initialized by the ASR-block with SpecAugment (last row of Table \ref{tab:sot}).
The setting (a) can be fairly compared with the LSTM-based SA-ASR model reported in \cite{kanda2020joint},
which was trained with the same data and used neither SpecAugment 
nor language models.
The transformer-based SA-ASR achieved an SA-WER of 8.2\%, outperforming the LSTM-based baseline by  47\% relative.
We 
observed a significant error reduction especially for the 2-speaker-mixed and 3-speaker-mixed test cases.
Then, the application of SpecAugment significantly improved
the SA-WER from 8.2\% to 7.6\% (setting (b)).
Finally, the proposed speaker deduplication further improved the SA-WER to 7.0\%, 
with notable SER improvement for 
the 3-speaker-mixed test case (setting (c)).

We also evaluated the speaker counting accuracy 
as shown in Table \ref{tab:spk-count}.
The transformer SA-ASR showed significantly better speaker counting accuracy than the LSTM-based counterpart.
By combining all techniques, the speaker counting accuracy finally reached  96--99\%
for all test conditions.
Note that the improvement from setting (a) to setting (c) was mostly attributed to the speaker dedupulication.

\subsection{Evaluation with LibriCSS}
\subsubsection{Experimental settings}
LibriCSS is 10 hours of audio recording made by playing back ``test\_clean" of LibriSpeech in a real meeting room.
Each recording consists of utterances from 8 speakers.
While the recording was conducted by a 7-ch microphone array,
we used only the first channel in this experiment.

We used the same training data set as the one  used in \cite{kanda2020investigation}.
The training data were generated by mixing 1 to 5 utterances of 1 to 5 speakers from LirbiSpeech,
where 
randomly generated room impulse responses and noise were added to simulate the reverberant recordings.
We initialized the ASR block by the model trained with SpecAugment,
and 
performed
160k iterations of training with a mini-batch of 6,000 frames with 8 GPUs.
Noam learning rate schedule with peak learning rate of 0.0001 after 10k iterations was used.
In addition to the SA-ASR model,
we trained a transformer-based language model (LM)
 (24-layers, 512-dim embedding, 8-heads MHA, 2048-dim feed forward) 
by using the same method in \cite{kanda2020investigation},
and used it based on  shallow fusion.

In the evaluation,
each long-form recording was first segmented at silence points detected by 
WebRTC Voice Activity Detector\footnote{https://github.com/wiseman/py-webrtcvad}
with the least aggressive setting.
Then, we ran the E2E SA-ASR with two different conditions.
In the first condition,
we ran the E2E SA-ASR model with relevant speaker profiles of 8 speakers in the meeting.
In the second condition,
we ran the combination of the E2E SA-ASR model and speaker clustering proposed in \cite{kanda2020investigation}. 
Specifically, the E2E SA-ASR model was executed with 100 dummy profiles,
and spectral clustering (SC) was applied on the speaker query of each utterance to assign  a speaker cluster tag. 
The number of speakers was given or estimated by the normalized maximum eigengap (NME) method \cite{park2019auto}.
The concatenated minimum-permutation word error rate (cpWER) \cite{watanabe2020chime} was used for the evaluation
to capture both the ASR and speaker diarization errors.

\subsubsection{Evaluation results}

The evaluation results are presented in Table \ref{tab:auto_summary}.
The proposed transformer-based E2E SA-ASR achieved a cpWER of 11.9\%.
This result was even better than
 that of the combination of 7-ch speech separation, spectral clustering-based speaker diarization and the transformer-based ASR (12.7\% cpWER) reported in \cite{raj2020integration}
 while the E2E SA-ASR used only a 1-ch signal. 

 The combination of the transformer SA-ASR and speaker clustering achieved 13.3\% or 16.3\% of cpWER with 
the oracle number of speakers or the estimated number of speakers, respectively.
These numbers are  significantly better than the result of the combination of the target-speaker voice activity detection (TS-VAD) and 
the transformer-based ASR (23.9\% of cpWER) reported in \cite{raj2020integration}
as well as LSTM-based SA-ASR \cite{kanda2020investigation}.

\section{Conclusions}

In this paper, we proposed the transformer-based E2E SA-ASR with 
the speaker deduplication mechanism.
The proposed model was demonstrated to be especially good at
counting speakers 
and
showed significant accuracy improvement for both the LibriSpeechMix and LibriCSS datasets.

\bibliographystyle{IEEEtran}

\bibliography{mybib}

\end{document}